\documentclass[twoside]{dis08}
\usepackage[latin1]{inputenc}
\usepackage[dvips]{graphicx,epsfig,color}
\usepackage{wrapfig,rotating}
\usepackage{amssymb,amsmath,array}

\begin{document}
\title{Update of the global fit of PDFs including the low-\boldmath{$Q$} 
DIS data}

\author{S.~Alekhin$^1$, S.~Kulagin$^2$, R.~Petti$^3$
\thanks{\uppercase{T}his work is partially supported by by the
\uppercase{RFBR} grant 06-02-16659.}
\vspace{.3cm}\\
1- Institute for High Energy Physics \\
142281 Protvino, Moscow region - Russia
\vspace{.1cm}\\
2- Institute for Nuclear researches \\
117312 Moscow - Russia\\
3- South Carolina University - Department of Physics and Astronomy\\
Columbia SC 29208 - USA\\
}

\maketitle

\begin{abstract}
We perform the next-to-next-leading-order (NNLO) QCD global fit of PDFs
using
inclusive charged-lepton and neutrino DIS data down to $Q = 1$ GeV.
We also consider
the data on neutrino-nucleon dimuon production, that allows us to disentangle
the strange sea distribution.
The fit results in $\chi^2/$NDP = 5150/4338 = 1.2 that demonstrates a
good consistency
of the data sets used in analysis.
The resulting value of
$\alpha_s(M_{\rm Z}) = 0.1136\pm 0.0007(exp.)$
 is in a good agreement with the previous version of the fit
with a more stringent cut on $Q$.
This analysis allows us to improve the accuracy of PDFs.
The HT terms of the neutrino-nucleon structure functions $F_2$ and
$xF_3$ are determined,
the former is found to be consistent with one for the charged-leptons
if the charge factor is taken into account.

\end{abstract}

\begin{wrapfigure}{r}{0.5\columnwidth}
\centerline{\includegraphics[width=0.45\columnwidth]{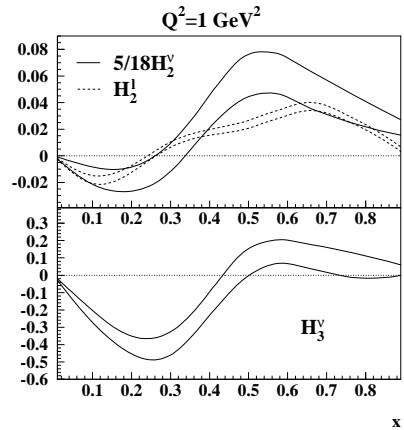}}
\caption{Upper panel: The $1\sigma$ band of the twist-4 term in the 
neutrino-nucleon SF $F_2$ (solid lines) compared to one for the 
charged lepton DIS (dashes). Lower panel: The same for the neutrino-nucleon 
SF $xF_3$.
}
\label{fig:htnu}
\end{wrapfigure}
A check of the existing deep-inelastic scattering (DIS)
 data at low values of the transferred momentum $Q$ is 
necessary ingredient of a parton distribution functions
(PDFs) fit. In principle, the concept of PDFs is applicable only 
at asymptotically big values of $Q$, 
where the QCD factorization is proved, while in the 
limit of small $Q$ it breaks 
due to the power corrections, which include the target mass effects 
and the dynamical high-twist (HT) contribution. 
Poor theoretical understanding of the latter does not allow to 
estimate the region of $Q$, where the power corrections can be neglected,
from the first principles; 
only phenomenological studies can separate effect of the HT terms. 
Earlier we have used the low-$Q$ DIS charged-leptons data
in the global fit of parton distribution functions
(PDFs) and found that these data 
can be well described in the NNLO of pQCD with account of the 
target mass corrections and additional
twist-4 terms down to $Q=1~{\rm GeV}$ \cite{Alekhin:2007zz}. 
The twist-4 terms in the structure functions (SFs) $F_2$ and $F_{\rm T}$ 
parameterized in this fit independently, 
by the smooth functions of general form, agree within the experimental errors.

For further study of the HT terms in the DIS SFs
we add to the fit of Ref.\cite{Alekhin:2007zz} the inclusive
(anti)neutrino-nucleon DIS data 
by the CHOURUS collaboration \cite{Onengut:2005kv}, 
down to $Q=1~{\rm GeV}$ as well. 
For the case of neutrino-nucleon DIS we have to take into account
3 additional independent HT terms, $H_2^{\nu}$, 
$H_{\rm T}^{\nu}$, and $H_3^{\nu}$ for the neutrino-nucleon SFs 
$F_2$, $F_{\rm T}$, and $xF_3$, respectively. 
These three terms cannot be disentangled using only the CHORUS data therefore
we impose constraint $H_2^{\nu}=H_{\rm T}^{\nu}$ motivated by the 
results obtained for the twist-4 terms in the charged-leptons SFs. 
We also assume that the HT terms are equal for 
the neutrino and anti-neutrino SFs. The $x$-dependence  
of $H_2^{\nu}$ and $H_3^{\nu}$
obtained under these assumptions is given in Fig.\ref{fig:htnu}. 
The former is in a good agreement with $H_2^{l}$ obtained for
the charged-leptons SFs scaled with the charge factor of 5/18.
This remarkable regularity is in favor of the dynamical nature of these terms.
In practice, this can also
be used to evaluate the HT terms for SFs for the neutral
current neutrino-nucleon DIS, which are experimentally poorly known.
The value of $H_3$ is basically negative, its integral over $x$ is 
$-0.084\pm0.033/(Q^2 [{\rm GeV}^2])$ that is in agreement with the early 
theoretical estimate of Ref.\cite{Braun:1986ty}.  
Earlier $H_3^{\nu}$ was also determined phenomenologically, 
in the NNLO QCD fit of Ref.\cite{Kataev:2001kk}.
Since that fit is based only on the data on $xF_3$  
the errors in $H_3^{\nu}$ obtained in Ref.\cite{Kataev:2001kk}
are quite big therefore comparison with those results 
is inconclusive. 

\begin{wrapfigure}{r}{0.5\columnwidth}
\centerline{\includegraphics[width=0.45\columnwidth]{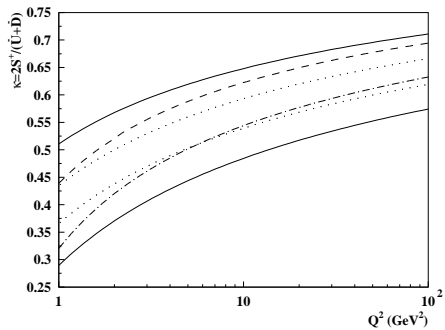}}
\caption{Upper panel: The $1\sigma$ band of the strange suppression factor 
$\kappa$ obtained in the fit with the value of semileptonic 
branching ratio released (solid lines) and fixed (dots). The value of
$\kappa$ for the CTEQ6 (dashed-dots) 
and MSTW06 (dashes) PDFs sets is given for comparison.
}
\label{fig:ssup}
\end{wrapfigure}

We also check impact of the data on dimuon production in the 
neutrino-nucleon scattering by the CCFR and NuTeV 
collaborations\cite{Mason:2006qa}
adding these data to the fit of Ref.\cite{Alekhin:2006zm}.
This input allows to constrain the strange quark distribution  
in the interval of $x=0.01\div 0.3$ that improves separation of 
the sea quarks distribution by flavors and eventually reduces the errors in PDFs
extracted from the global fit. 
The $O(\alpha_{\rm s})$ (NLO QCD) corrections of Ref.\cite{Gottschalk:1980rv}
to the neutrino-nucleon charm production were taken into account. 
The HT contribution to the dimuon cross section was not included.
Once we try to add the HT terms,
which are observed in the inclusive SFs with account of the 
corresponding charge factors, results of the fit do not change. 
For this reason we include into the fit dimuon data with $Q$ down to 
$1~{\rm GeV}$, as in the case of inclusive cross sections.  
Traditionally, magnitude of the strange sea is described 
by the strange sea suppression factor $\kappa$ equal to the
ratio of momentum carried by the strange quarks to the non-strange one. 
The value of $\kappa$ obtained in our fit with 
the dimuon CCFR/NuTeV data included is given in Fig.\ref{fig:ssup}.
This value is bigger than ones obtained in the NLO QCD fit of 
Ref.\cite{Bazarko:1994tt} based on the CCFR data and in the LO fit
of Ref.\cite{Goncharov:2001qe} 
based on the CCFR/NuTeV data (0.48 and some 0.42
at $Q^2=20~{\rm GeV}^2$, respectively).
\begin{wrapfigure}{r}{0.5\columnwidth}
\centerline{\includegraphics[width=0.45\columnwidth]{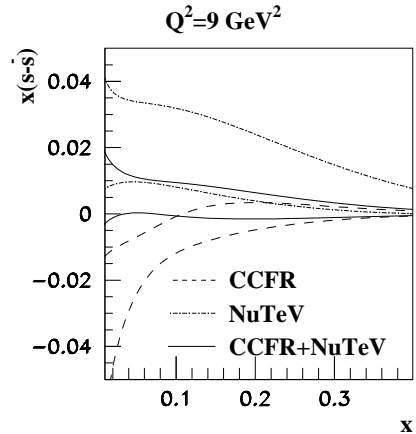}}
\caption{The $1\sigma$ band for the strange/antistrange sea asymmetry 
obtained using the NuTeV (dashed dots), CCFR (dashes) and combination of 
both (solid lines) dimuon data sets. 
}
\label{fig:ssminus}
\end{wrapfigure}
Tracing this difference we found that it stems due to the non-strange sea 
in the PDFs set BGPAR used in Refs.\cite{Bazarko:1994tt,Goncharov:2001qe} 
is enhanced as compared to our results and other modern PDFs parameterizations.
For the LO fit of Ref.\cite{Goncharov:2001qe} the difference with our 
result is more significant since account of the NLO corrections leads
to increase of the fitted magnitude of the strange sea. 
The values of $\kappa$ calculated using the CTEQ6\cite{Pumplin:2002vw} and 
MSTW06\cite{Martin:2007bv} PDFs sets are in agreement to our determination, 
the former is some smaller and the latter is some bigger than ours,  
but in both cases the difference is within the uncertainty in $\kappa$.

The dominant source of this uncertainty is due to the error in
the charm quark semileptonic branching ratio $B_c$. 
The value of $B_c$ depends on the charmed hadrons production fractions 
in the neutrino-nucleon scattering, which are not known very well and, through 
these fractions, depends on the beam energy. 
For the beam energy typical for the CCFR and NuTeV experiments the 
value of $B_{\rm c}$ was estimated in Ref.\cite{Bolton:1997pq}
using the charm fractions  by the Fermilab  
experiment E531~\cite{Ushida:1988rt} and 
the charmed hadrons branching ratios measured in the $e^+e^-$ experiments. 
With updated values of the charmed hadrons branching ratios \cite{BR} 
we obtain the value $B_{\rm c}=0.0886(57)$. We fit $B_{\rm c}$ to the dimuon data 
simultaneously with the strange sea and we get 
a value of $0.089(7)$, which is in a good agreement to both estimates. 
The beam energy dependence of $B_{\rm c}$ in the kinematic region covered 
by the CCFR/NuTeV data is found to be inessential for our fit. 
The fitted slope of $B_{\rm c}$ on the beam energy
is well consistent with zero within the errors.   

The value of strange/anti-strange (s/sbar) sea symmetry we obtain is 
given in Fig.\ref{fig:ssminus}. If we include the 
NuTeV dimuon data only it is in agreement 
to one determined by the NuTeV collaboration from the NLO QCD
analysis of their data\cite{Mason:2007zz}. 
The value of this asymmetry we obtain using the CCFR data only is 
also qualitatively consistent with one observed in the LO QCD fit 
of Ref.\cite{Goncharov:2001qe} (impact of the QCD correction on the 
asymmetry is marginal and does not affect the comparison). 
Averaging of these two results gives the asymmetry consistent with zero 
since the NuTeV and CCFR determinations are comparable in magnitude, but 
have different sign. 
Despite we do not set the constraint on the net strangeness in the
nucleon, the resulting integral of s/sbar asymmetry is 0.0011(13), i.e.
comparable to zero; in the
variant of fit with the net starngeness set to 0, the value of
$\chi^2$ gets bigger nothing but
by 1. This disagrees with the results of two other 
groups\cite{Lai:2007dq,MSTW}, who observe
positive s/sbar asymmetry at $x \sim 0.2$ with statistical
significance of about $2\sigma$.
In this context
one can note that the resulting s/sbar asymmetry is rather small, and,
therefore,
sensitive to subtle details of the fit: the specific value of $B_c$,
the treatment of nuclear corrections, the choice of the cut on $Q$.
Each of these factors can change the resulting value of s/sbar by about 0.01,
and the combinations of them, in principle, can explain the difference
with the results of Refs.\cite{Lai:2007dq,MSTW}.

Combining the inclusive CHORUS data and the CCFR/NuTeV dimuon data with the
data set used in the analysis of Ref.\cite{Alekhin:2007zz}, 
we perform the global fit of PDF with the
extraction of the HT terms in the charged-leptons and neutrino-nucleon
inclusive DIS.
In this fit, the
total value of $\chi^2/$NDP = 5150/4338 = 1.2, which is somewhat
bigger than the ideal
value of 1. However, if we rescale the errors in data of the separate
experiments, which have
$\chi^2/{\rm NDP} > 1$, in order to bring them to 1, and then recalculate
the errors in PDFs, the
change in these errors is moderate, factor of 1.2 at most. Thus the data are in
satisfactory consistency with no need of the big overall rescaling of
the errors for
all data sets. The PDF accuracy improved as compared to PDFs of
Ref.\cite{Alekhin:2007zz}, because of
additional experimental input. The typical errors in PDFs are $O(1\%)$
at $x \lesssim 0.1$.
At larger
values of $x$ the errors are bigger with the gluon distribution
especially uncertain. The
extracted value of $\alpha_s(M_{\rm Z}) = 0.1136\pm 0.0007(exp.)$ is in a
good agreement with the
result of the fit of Ref.\cite{Alekhin:2007zz}, 
in which a more stringent cut on Q was applied.
This justifies the use of the NNLO QCD analysis combined with the
twist expansion down to $Q = 1$~GeV.
Therefore, the PDFs, extracted in our fit, are relevant for the
low-energy studies.
A small experimental error in $\alpha_s$ in our fit is because of the
impact of the low-$Q$ data, 
which are very sensitive to the details of the QCD evolution.
On the other hand, the theoretical error rises at small values of $Q$,
due to the variation of QCD scales.
This becomes the dominant source of the uncertainty in
$\alpha_s$, and, in order to reduce it, a consistent account of higher-order QCD
corrections is necessary.

\begin{footnotesize}

\end{footnotesize}

\begin{thebibliography}{99}

\bibitem{url} Slides: \\
\verb$http://indico.cern.ch/contributionDisplay.py?contribId=181&sessionId=17&confId=24657$

\bibitem{Alekhin:2007zz}
  S.~Alekhin, S.~Kulagin and R.~Petti,
{\it Prepared for 15th International Workshop on Deep-Inelastic Scattering and Related Subjects (DIS2007), Munich, Germany, 16-20 Apr 2007}

\bibitem{Onengut:2005kv}
  G.~Onengut {\it et al.}  [CHORUS Collaboration],
  Phys.\ Lett.\  B {\bf 632} (2006) 65.

\bibitem{Braun:1986ty}
  V.~M.~Braun and A.~V.~Kolesnichenko,
  Nucl.\ Phys.\  B {\bf 283} (1987) 723.

\bibitem{Kataev:2001kk}
  A.~L.~Kataev, G.~Parente and A.~V.~Sidorov,
  Phys.\ Part.\ Nucl.\  {\bf 34}, 20 (2003)
  [Fiz.\ Elem.\ Chast.\ Atom.\ Yadra {\bf 34}, 43 (2003\ ERRAT,38,827-827.2007)]

\bibitem{Mason:2006qa}
  D.~A.~Mason, FERMILAB-THESIS-2006-01, UMI-32-11223, 2006. 

\bibitem{Alekhin:2006zm}
  S.~Alekhin, K.~Melnikov and F.~Petriello,
  Phys.\ Rev.\  D {\bf 74}, 054033 (2006).

\bibitem{Gottschalk:1980rv}
  T.~Gottschalk,
  Phys.\ Rev.\  D {\bf 23} (1981) 56.

\bibitem{Bazarko:1994tt}
  A.~O.~Bazarko {\it et al.}  [CCFR Collaboration],
  Z.\ Phys.\  C {\bf 65} (1995) 189

\bibitem{Goncharov:2001qe}
  M.~Goncharov {\it et al.}  [NuTeV Collaboration],
  Phys.\ Rev.\  D {\bf 64} (2001) 112006.

\bibitem{Pumplin:2002vw}
  J.~Pumplin, D.~R.~Stump, J.~Huston, H.~L.~Lai, P.~Nadolsky and W.~K.~Tung,
  JHEP {\bf 0207}, 012 (2002).

\bibitem{Martin:2007bv}
  A.~D.~Martin, W.~J.~Stirling, R.~S.~Thorne and G.~Watt,
  Phys.\ Lett.\  B {\bf 652}, 292 (2007)

\bibitem{Bolton:1997pq}
  T.~Bolton,
  arXiv:hep-ex/9708014.

\bibitem{Ushida:1988rt}
  N.~Ushida {\it et al.}  [Fermilab E531 Collaboration],
  Phys.\ Lett.\  B {\bf 206} (1988) 375.

\bibitem{BR}
  N.~E.~Adam {\it et al.}  [CLEO Collaboration],
  Phys.\ Rev.\ Lett.\  {\bf 97} (2006) 251801;
  T.~K.~Pedlar {\it et al.}  [CLEO Collaboration],
  Phys.\ Rev.\  D {\bf 76} (2007) 072002;
  K.~M.~Ecklund {\it et al.}  [CLEO Collaboration],
  Phys.\ Rev.\ Lett.\  {\bf 100} (2008) 161801;

\bibitem{Mason:2007zz}
  D.~Mason {\it et al.},
  Phys.\ Rev.\ Lett.\  {\bf 99} (2007) 192001.

\bibitem{Lai:2007dq}
  H.~L.~Lai, P.~Nadolsky, J.~Pumplin, D.~Stump, W.~K.~Tung and C.~P.~Yuan,
  JHEP {\bf 0704}, 089 (2007)
\bibitem{MSTW}
  G.~Watt, A.~D.~Martin, W.~J.~Stirling and R.~S.~Thorne,
  arXiv:0806.4890 [hep-ph].

\end{thebibliography}
\end{document}